\documentclass[11pt,floatfix,aps,prl,amsmath,nofootinbib,preprintnumbers,onecolumn,preprint]{revtex4}
\usepackage{graphicx}
\usepackage{epstopdf}
\usepackage{amsmath}
\usepackage{amsfonts}
\usepackage{amssymb}
\usepackage{color}
\usepackage{mathrsfs}
\usepackage{multirow}
\usepackage{bm}
\usepackage{subfigure}
\usepackage{xcolor}
\usepackage{mathtext,bbm,amsmath,amsfonts,amssymb,indentfirst,syntonly,graphicx}
\usepackage{mathtools}
\usepackage{slashbox}
\usepackage[english]{babel}
\usepackage{calc}
\usepackage{tikz}
\usepackage[normalem]{ulem}

\def\({\left(}
\def\){\right)}
\def\[{\left[}
\def\]{\right]}
\def\e{\begin{equation}}
\def\q{\end{equation}}
\def\m{\begin{eqnarray}}
\def\n{\end{eqnarray}}

\begin{document}

\title{Constraining the Lorentz invariance violation from the continuous spectra of short gamma-ray bursts}

\author{Zhe Chang $^{1}$}

\author{Xin Li $^{2}$}

\author{Hai-Nan Lin $^{2}$}

\author{Yu Sang $^1$}
\thanks{Corresponding author at: sangyu@ihep.ac.cn}

\author{Ping Wang $^{1}$}

\author{Sai Wang $^{3}$}

\affiliation{$^1$Institute of High Energy Physics, Chinese Academy of Sciences, Beijing 100049, China\\
$^2$Department of Physics, Chongqing University, Chongqing 401331, China\\
$^3$State Key Laboratory of Theoretical Physics, Institute of Theoretical Physics, Chinese Academy of Sciences, Beijing 100190, China}

\date{\today}

\begin{abstract}
In some quantum gravity theories, a foamy structure of space-time may lead to Lorentz invariance violation (LIV). As the most energetic explosions in the Universe, gamma-ray bursts (GRBs) provide an effect way to probe quantum gravity effects. In this paper, we use the continuous spectra of 20 short GRBs detected by the Swift satellite to give a conservative lower limit of quantum gravity energy scale $M_\textrm{QG} $. Due to the LIV effect, photons with different energy have different velocities. This will lead to the delayed arrival of high energy photons relative to low energy ones. Based on the fact that the LIV-induced time delay cannot be longer than the duration of a GRB, we present the most conservative estimate of the quantum gravity energy scales from 20 short GRBs. The strictest constraint, $M_\textrm{QG}>5.05\times10^{14}$ GeV in the linearly corrected case, is from GRB 140622A. Our constraint on $M_{\rm QG}$, although not as tight as previous results, is the safest and most reliable so far.
\end{abstract}

\maketitle


\section{I. Introduction}\label{sec:introduction}

A foamy structure of space-time on short time and distance scales has been proposed in quantum gravity \cite{Amelino-Camelia:1997,Ellis:2008a,Ellis:2011}. The non-trivial space-time can lead to the violation of Lorentz invariance at the Planck scale. Probing the Lorentz invariance violation (LIV) effects provides a useful way to test the validity of quantum gravity theories, such as loop quantum gravity \cite{Gambini:1999,Alfaro:2002}, string theory \cite{Ellis:1992,Ellis:1999} and double special relativity \cite{Amelino-Camelia:2002}. According to the LIV effect, massless particles have energy-dependent velocities. Hence, the velocity of a photon propagating in vacuum may have a tiny deviation from the trivial value, $c$. The delay of arrival time induced by the LIV effect is a monotonically increasing function of the photon energy and the distance of source. We need plenty of distant energetic photons in order to observe the LIV effect.

As the most energetic explosions in the Universe, gamma-ray bursts (GRBs) can be detectable up to distances as far as tens of Gpc away from us. It has been proposed that GRBs provide an effective way to probe the LIV effect because of their cosmological distance and rapid emissions of energetic photons \cite{Amelino-Camelia:1998,Ellis:2013}. In fact, GRBs have already been widely used to constrain the LIV effect \cite{Chang:2012,Shao:2010,Zhang:2014wpb,Amelino-Camelia:1998, Vasileiou:2013,Abdo:2009,Ellis:2006,Vasileiou:2015,Nemiroff:2012, Ellis:2003,Boggs:2004,Zhang:2015, Pan:2015,Ellis:2013,Xiao:2009,Kosteleck:2013, Kahniashvili:2006,Coleman:1999,Biesiada:2009, Gogberashvili:2007,Schaefer:1999}. However, since we have little knowledge of the emission mechanism of GRBs, we cannot distinguish the LIV-induced time delay from the intrinsic time delay. Ellis et al. \cite{Ellis:2006} performed a linear regression analysis of GRBs with measured redshift. The data were fitted by a straight line with a slope corresponding to the quantum gravity scale and the intercept representing the possible intrinsic time delay inherited from the sources. They found a strong correlation between the parameters characterizing an intrinsic time delay and a distance-dependent propagation effect. Their work based on the assumption that all GRBs had the same intrinsic emission mechanism and the intrinsic time delay. Since the durations of GRBs span about 6 orders of magnitude, there is no persuasive reason to believe that two high energy photons from two different GRBs (or two photons with different energy from the same GRB) have the same intrinsic time delay relative to the trigger time of low energy photons. As an improvement, Zhang and Ma \cite{Zhang:2015} fitted the data with straight lines of the same slope but different intercepts. However, photons on the same line must have the same intrinsic time delay. This is not always true, because photons on the same line may come from different GRBs and have very different energy. Chang et al. \cite{Chang:2012} used the magnetic jet model to estimate the intrinsic time delay between emissions of low and high energy photons. Unfortunately, the magnetic jet model depends on some unobservable parameters, and thus introduces large uncertainties. Vasileiou et al. \cite{Vasileiou:2015} adopted maximum-likelihood analysis to test quantum gravity model in which photons had normally distributed velocities. They used the synthetic data of GRB 090510, and chose the time interval and threshold energy $E_\textrm{th}$ according to increased sensitivity and minimal systematic biases. However, in their calculation, only hundreds of photons below $E_\textrm{th}$ and tens above $E_\textrm{th}$ were used.

The previous works can be divided into two major classes. The first class constrains $M_{\rm QG}$ using one or some isolated high energy photons from GRBs \cite{Shao:2010,Chang:2012,Zhang:2014wpb}. The second class constrains $M_{\rm QG}$ using the observed spectral lag between the low and high energy band \cite{Ellis:2003,Boggs:2004,Ellis:2006,Abdo:2009,Nemiroff:2012,Vasileiou:2013,Vasileiou:2015}. The first method is not statistically significant because the number of high energy photons is too small. More importantly, we cannot know if high energy photons and low energy ones are emitted simultaneously. The second method is based on the fact that, in some Fermi GRBs, high energy photons show a systematic spectral lag with respect to low energy photons. By assuming that the LIV-induced time delay cannot be longer than the spectral lag, $M_{\rm QG}$ can be constrained to the order of Planck energy scale. This method has the underling assumption that the low and high energy photons are emitted in the same region, which is widely accepted to be true since low and high energy photons form the same types of spectra (such as the Band function). However, there is still a possibility that low and high energy photons come from different regions. For example, Kumar \& Duran \cite{Kumar:2009ps} showed that the high-energy photons ($\gtrsim 100$ MeV) of GRB 080916C were unambiguously generated in the external shock via the synchrotron process, while the lower energy photons had a distinctly different source. Besides, Ellis et al. \cite{Ellis:2006} analyzed the light curves of 35 GRBs and found a systematic tendency for more energetic photons to arrive earlier than low energy photons. This conflicts with most Fermi GRBs which often show that high energy photons come later than low energy ones \cite{Abdo:2009pg,Ackermann:2010us,Ackermann:2011bu}. With the present knowledge, we cannot distinguish the LIV-induced time delay and the intrinsic time delay from the observed spectral lag. If high energy photons are intrinsically emitted earlier than low energy photons (we can't exclude this possibility even high energy photons are observed later), the LIV-induced time delay may be much longer than the observed spectral lag. On the other hand, most short GRBs show no significant spectral lag. If we still use the spectral lag as the upper limit of LIV-induced time delay, we may constrain $M_{\rm QG}$ to be infinity. But we cannot safely say that the LIV effect does not exist.

In this paper, for the first time, we use the duration of short GRBs as the upper limit of the LIV-induced time delay. This is always true, regardless of whether the observed spectral lag is totally due to the intrinsic or LIV-induced time delay (or the combined contribution of these two). We use 20 short GRBs with measured redshift from Swift \cite{SWIFT} to give a conservative lower limit of quantum gravity energy scale. Whether the intrinsic emission process contributes to the time delay or not, the LIV-induced time delay is certainly smaller than the duration of a GRB. The rest of this paper is arranged as follows. In Section II, we review the observational properties of GRBs. In Section III, quantum gravity energy scale is constrained using short GRBs. Finally, a short summary is given in Section IV.

\section{II. Observational properties of GRBs}\label{sec:observations}

Observations of GRBs, especially by means of the Compton, Swift, and Fermi satellites, contribute to the research of their properties. One of the most important properties of GRBs is the duration. The duration of a GRB is usually characterized by $T_{90}$, during which from 5\% to 95\% of the total photon events in a specific energy band are detected. However, the duration $T_{90}$ depends on the sensitivity of the detector and the energy band in which the detector works. A detector with a lower and broader energy band generally gets a longer $T_{90}$ for the same GRB \cite{Kumar:2015}. The observed durations span about 6 orders of magnitude, from milliseconds to thousands of seconds. The distribution of durations is bimodal and separated at around 2 s. Thus, Kouveliotou et al. \cite{Kouveliotou:1993} proposed a classification according to $T_{90}$: long bursts with $T_{90}>2$ s and short bursts with $T_{90}<2$ s. The GRB light curves are irregular. Some are variable with many peaks, while some are smooth with simple temporal structures.

The GRB spectra are non-thermal. The energy of the prompt emission is concentrated in the hundreds of keV range. In some of the brightest GRBs, photons with energy higher then 100 MeV (and maybe up to tens of GeV) have been observed \cite{Abdo:2009,Abdo:2009pg,Ackermann:2010us,Ackermann:2011bu}. X-ray emission is weak, and a small part of the emissions are below 10 keV \cite{Piran:1999}. A typical GRB spectrum can be fitted with the so-called Band function \cite{Band:1993}. It can be written as
\begin{eqnarray}\label{Band function}
N(E)=
  \begin{cases}
  \displaystyle A\left(\frac{E}{100~\rm{keV}}\right)^\alpha \exp\left(-\frac{E}{E_0}\right), &  E<\left(\alpha-\beta\right)E_0 \\
  \displaystyle A\left[\frac{\left(\alpha-\beta\right)E_0}{100~\rm{keV}}\right]^{\alpha-\beta} \exp\left(\beta-\alpha\right)\left(\frac{E}{100~\rm{keV}}\right)^\beta, & E\geq\left(\alpha-\beta\right)E_0\\
  \end{cases}
\end{eqnarray}
where $N(E)$\rm{d}$E$ is the photon number in energy interval d$E$, $\alpha$ and $\beta$ are the spectral indices, and $E_0$ is the break energy. Typical evaluations for the spectral parameters are $\alpha\approx -0.5 \sim -1.5$, $\beta\approx -2 \sim -3$, $E_0\approx 0.1 \sim 1$ MeV \cite{Preece:1998,Lloyd:2000}. Eq.(\ref{Band function}) is a time integrated spectrum of photon number and consists of two smoothly-joined power law parts. The Band function peaks at around a few hundred keV and it fits well to most of the observed spectra \cite{Kumar:2015}.

The Swift satellite, launched in November 2004, has three instruments working together to observe GRBs in prompt emission and afterglow phases in the gamma-ray, X-ray, ultraviolet, and optical wavebands. The Burst Alert Telescope (BAT) first detects the GRB and accurately determines its direction in the sky. Then in less than approximately 90 seconds, the X-ray Telescope (XRT) and UV-Optical Telescope (UVOT) slew to the GRB and start observing. Swift provides the light curves of prompt emission and afterglows and evaluates the duration $T_{90}$ of a GRB. In the following, we use photons detected by BAT, which works in the $15-150$ keV energy band.

\section{III. Constraints on quantum gravity energy scale}\label{sec:constraint}

In quantum gravity, there are several possible scenarios with respect to the breaking of Lorentz invariance. In consideration of experimental tests, Amelino-Camelia and Smolin \cite{Amelino-Camelia:2009} sorted them into three broad categories, i.e., naive Lorentz symmetry breaking, Lorentz symmetry breaking in effective field theory, and doubly special relativity. In all three models, Lorentz symmetry is broken at a very high energy scale, which is expected to be around the Planck energy.  In general, the deformed dispersion relation of massless particles at leading term can be written as \cite{Jacob:2008}
\begin{equation}\label{dispersion relation}
E^2\simeq p^2c^2\left[1\pm\left(\frac{pc}{M_{\textrm{QG}}}\right)^n\right],
\end{equation}
where $n = 1,2,3$ denotes the linear, quadratic and cubic corrections to the dispersion relation, respectively, and $M_\textrm{QG}$ represents the quantum gravity energy scale. The `+' or `--' in Eq.(\ref{dispersion relation}) corresponds to the superluminal or subluminal motion of particles. When the energy of a particle reaches a value at the same order of $M_\textrm{QG} $, the LIV effects will become obvious. We use the `--' case where particles with high energies travel slower than those with low energies.

As a result of the LIV effect, two photons with different energies, emitted simultaneously from the same GRB, have different travel speeds and arrive at the earth at different times. Due to the cosmological distance and the high energy of GRB photons, the time delay is accumulated as the propagation of photons and should be measurable. Taking into consideration the expansion of the Universe, we get the expression of the LIV-induced time delay of two GRB photons \cite{Jacob:2008,Ellis:2008}:
\begin{equation}\label{LIV time delay}
\Delta t_\textrm{LIV}=\frac{1+n}{2H_0}\left(\frac{\Delta E}{M_\textrm{QG}}\right)^n\int_0^z dz'\frac{(1+z')^n}{\sqrt{\Omega_m\left(1+z'\right)^3+\Omega_\Lambda}},
\end{equation}
where $\Delta E$ is the energy difference of two photons, and $z$ is the redshift of the GRB source. $H_0$ is the Hubble constant, and $\Omega_m$ and $\Omega_\Lambda$ are the present values of the matter density and cosmological constant density, respectively. Throughout this paper, the Planck 2015 results are used, i.e., $H_0=67.8~\rm{km~s^{-1}~Mpc^{-1}}$, $\Omega_m=0.308$, and $\Omega_\Lambda=0.692$ \cite{Ade:2015}.

\begin{figure}[htbp]
  \centering
 \includegraphics[width=10 cm]{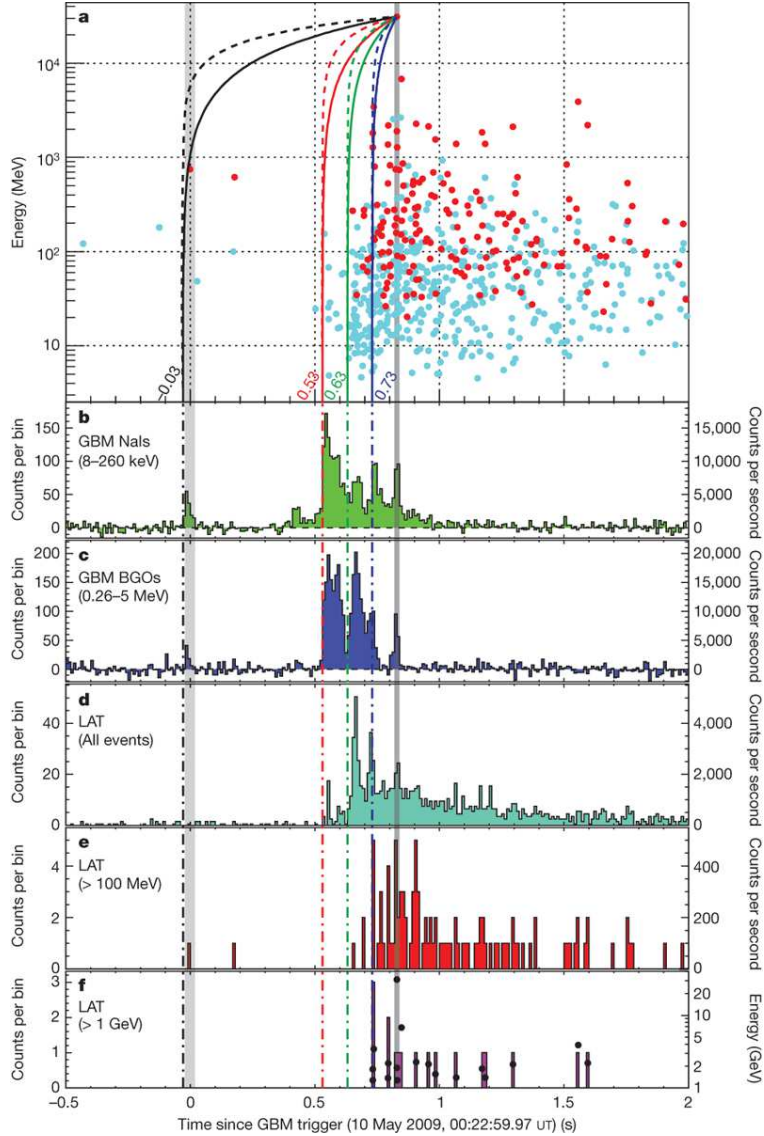}
 \caption{\small{The light curves of GRB 090510 in different energy bands (figure reproduced from reference \cite{Abdo:2009}). Panel(a): LAT photons passing the off-line (red) and onboard (blue) event selections. The lines represent the relation between photon energy and arrival time for linear (solid lines) and quadratic (dashed lines) LIV. Panel(b-f): GBM and LAT light curves, from lowest to highest energies. Panel(f) shows the isolated photons with energy $E>1$ GeV.}}\label{fig:090510}
\end{figure}

As was mentioned in the introduction, some works constrained LIV using a small number of GeV photons selected from a few GRBs. The energy spectrum of a GRB at the GeV scale is usually discrete. Only a few GeV photons can be detected in a GRB emission \cite{Rubtsov:2012}. The results of using GeV photons to constrain LIV are unpersuasive in statistics. However, at the keV order, the energy spectrum is continuous, and thousands of photon events can be recorded. Thus, using keV photons from GRBs to constrain LIV is more statistically reliable. Taking GRB 090510, for example, Figure \ref{fig:090510} shows the light curves of GRB 090510 at different energy bands detected by the Fermi satellite.
Panel (b) is the relation of events and time in the 8 -- 260 keV energy band, and the maximum number of events is up to 17,000 counts per second. In Panel (f) there are only 18 photons with $E > 1~\textrm{GeV}$ detected by Fermi-LAT.

We analyze GRBs detected by the Burst Alert Telescope (BAT) onboard the Swift satellite, and concentrate on the 90\% of photons in an energy band instead of single photons. BAT can catch photons in the energy band of $15 - 150$ keV, and record the durations of GRBs, $T_{90}$. It is certain that the LIV-induced time delay ($\Delta t_{\rm LIV}$) cannot be longer than the duration of a GRB. Otherwise, the LIV effect would spread the duration of the GRB until it is longer than $\Delta t_{\rm LIV}$. Therefore, $T_{90}$ is the upper limit of the LIV-induced time delay. Thus we always have
\begin{equation}\label{uplimit}
  \Delta t_{\rm LIV}<T_{90}.
\end{equation}
Eq. (\ref{uplimit}) is a much more conservative but safer estimation of $\Delta t_{\rm LIV}$ compared to previous works using spectral lag as the upper limit. Substituting Eq. (\ref{LIV time delay}) into Eq. (\ref{uplimit}), we finally get
\begin{equation}\label{M_QG}
M_\textrm{QG}>\Delta E\left[\frac{(1+n)F_n(z)}{2H_0T_{90}}\right]^{\frac{1}{n}},
\end{equation}
where
\begin{equation}
  F_n(z)\equiv \int_0^z dz'\frac{(1+z')^n}{\sqrt{\Omega_m\left(1+z'\right)^3+\Omega_\Lambda}}.
\end{equation}
From Eq. (\ref{M_QG}), we can see that a rigorous limit on $M_\textrm{QG}$ requires GRBs of short duration. We choose GRBs from Swift data archive, and only short GRBs with duration $T_{90} < 2$ s are selected. Our sample consists of 20 short GRBs with measured redshift in the range $z\in[0.093,2.609]$. The GRB sample is listed in Table \ref{tab:results}.
\begin{table}[htbp]
\caption{\small{The properties of 20 short Swift GRBs we used in the calculation and the resulting quantum gravity energy scales. $z$ is the redshift, $T_{90}$ is the duration, and $M_\textrm{QG}$ is the lower limit of quantum gravity energy scale in the linearly corrected case.}}\label{tab:results}
\begin{tabular}{lccccc}
\hline\hline
GRB ~~~~~~~~& ~~~~$z$~~~~ &~~~~ $T_{90}$~~~~ & $M_\textrm{QG}$ \\
& & (s) & $(10^{14}\rm{GeV})$ \\
\hline
150120A	&	0.46	&	1.2	&	0.26 	 	\\
150101B	&	0.093	&	0.018	&	3.25 	 	\\
141212A	&	0.596	&	0.3	&	1.34 		\\
140903A	&	0.351	&	0.3	&	0.77 	 	\\
140622A	&	0.959	&	0.13	&	5.05 	 	\\
131004A	&	0.717	&	1.54	&	0.32 	 	\\
130603B	&	0.356	&	0.18	&	1.30 	 	\\
101219A	&	0.718	&	0.6	&	0.81 		 	\\
100724A	&	1.288	&	1.4	&	0.63 		\\
090510	&	0.903	&	0.3	&	2.06 	 	\\
090426	&	2.609	&	1.2	&	1.42 		\\
071227	&	0.383	&	1.8	&	0.14 	  \\
070724A	&	0.457	&	0.4	&	0.76 	 	\\
070429B	&	0.904	&	0.47	&	1.32 	 	\\
061217	&	0.827	&	0.21	&	2.69 	 	\\
061201	&	0.111	&	0.76	&	0.09 		\\
060502B	&	0.287	&	0.131	&	1.43 	 	\\
051221A	&	0.547	&	1.4	&	0.26 		\\
050813	&	1.8	&	0.45	&	2.70 		\\
050509B	&	0.225	&	0.073	&	1.99 	 	\\
\hline
\end{tabular}
\end{table}
We use the duration $T_{90}$ as the upper limit of $\Delta t_\textrm{LIV}$ between lower energy (15 keV) photons and higher energy (150 keV) photons. Namely, we get a conservative lower limit of the quantum gravity energy scale $M_\textrm{QG} $.

Here we mainly focus on the linear correction case, i.e., the $n=1$ case. The results are listed in the fourth column of Table \ref{tab:results}, and we get a conservative limit $M_\textrm{QG}>5.05\times10^{14}$ GeV from GRB 140622A for linear correction of the dispersion relation. Most of the GRBs give an order of $10^{14}$ GeV, and the most rigorous constraint is given by GRB 140622A because of its high redshift and short duration. Although GRB 090426 has the highest redshift, its duration is longer than most other GRBs. Due to the low redshift and relatively long duration, GRB 061201 gives a limit of $M_\textrm{QG}$ only at the order of $\sim 10^{12}$ GeV. Our constraint on $M_\textrm{QG}$ is much looser than previous results \cite{Xiao:2009,Kosteleck:2013,Kahniashvili:2006,Coleman:1999,Nemiroff:2012, Biesiada:2009,Shao:2010, Gogberashvili:2007,Ellis:2003,Schaefer:1999,Boggs:2004}, because  the energy band we used is narrow. However, our result is much more statistically significant. This is because the whole energy band in $15-150$ keV is used in our calculation, while other works only chose a limited number of high-energy photons.

Finally, we also use GRB 140622A to constrain $M_{\rm QG}$ in the quadratic and cubic correction cases of the dispersion relation, i.e., the $n=2$ and $n=3$ cases in Eq.(\ref{dispersion relation}). We get a conservative limit $M_{\rm QG} > 3.90\times 10^5$ GeV for the quadratic correction, and $M_{\rm QG} > 3.48 \times 10^2$ GeV for the cubic correction. The constraint on the LIV effect in the quadratic and cubic cases is much looser than in the linear case.

\section{IV. Summary}\label{sec:conclusion}

In summary, we have used 20 short GRBs with redshift measurements from the Swift data archive to constrain the possible LIV effect that was predicted by some quantum gravity theories. Based on the fact that the LIV-induced time delay should not be longer than the duration of a GRB, we derived the lower limit of quantum gravity energy scale at the order of $M_\textrm{QG} \sim 10^{14}$ GeV. The strictest limit was given by GRB 140622A, i.e., $M_\textrm{QG}>5.05\times10^{14}$ GeV. Our constraint on $M_{\rm QG}$, although not as tight as previous results, is safer and more reliable than previous studies. GRBs with higher redshift and shorter duration give a tighter constraint on $M_\textrm{QG}$. Future observations at higher and broader energy bands will also tighten the constraint.

\begin{acknowledgments}
We are grateful to D. Zhao for useful discussions. This work has been supported by the National Natural Science Foundation of China under grant Nos. 11375203, 11305181, 11322545 and 11335012, and by Knowledge Innovation Program of The Chinese Academy of Sciences.
\end{acknowledgments}





\bibliographystyle{apsrev}
\bibliography{myreference}

\begin{thebibliography}{45}
\expandafter\ifx\csname natexlab\endcsname\relax\def\natexlab#1{#1}\fi
\expandafter\ifx\csname bibnamefont\endcsname\relax
  \def\bibnamefont#1{#1}\fi
\expandafter\ifx\csname bibfnamefont\endcsname\relax
  \def\bibfnamefont#1{#1}\fi
\expandafter\ifx\csname citenamefont\endcsname\relax
  \def\citenamefont#1{#1}\fi
\expandafter\ifx\csname url\endcsname\relax
  \def\url#1{\texttt{#1}}\fi
\expandafter\ifx\csname urlprefix\endcsname\relax\def\urlprefix{URL }\fi
\providecommand{\bibinfo}[2]{#2}
\providecommand{\eprint}[2][]{\url{#2}}

\bibitem[{\citenamefont{Amelino-Camelia
  et~al.}(1997)\citenamefont{Amelino-Camelia, Ellis, Mavromatos, and
  Nanopoulos}}]{Amelino-Camelia:1997}
\bibinfo{author}{\bibfnamefont{G.}~\bibnamefont{Amelino-Camelia}},
  \bibinfo{author}{\bibfnamefont{J.}~\bibnamefont{Ellis}},
  \bibinfo{author}{\bibfnamefont{N.~E.} \bibnamefont{Mavromatos}},
  \bibnamefont{and} \bibinfo{author}{\bibfnamefont{D.~V.}
  \bibnamefont{Nanopoulos}}, \bibinfo{journal}{Int. J. Mod. Phys. A}
  \textbf{\bibinfo{volume}{12}}, \bibinfo{pages}{607} (\bibinfo{year}{1997}).

\bibitem[{\citenamefont{Ellis et~al.}(2008{\natexlab{a}})\citenamefont{Ellis,
  Mavromatos, and Nanopoulos}}]{Ellis:2008a}
\bibinfo{author}{\bibfnamefont{J.}~\bibnamefont{Ellis}},
  \bibinfo{author}{\bibfnamefont{N.~E.} \bibnamefont{Mavromatos}},
  \bibnamefont{and} \bibinfo{author}{\bibfnamefont{D.~V.}
  \bibnamefont{Nanopoulos}}, \bibinfo{journal}{Phys. Lett. B}
  \textbf{\bibinfo{volume}{665}}, \bibinfo{pages}{412}
  (\bibinfo{year}{2008}{\natexlab{a}}).

\bibitem[{\citenamefont{Ellis et~al.}(2011)\citenamefont{Ellis, Mavromatos, and
  Nanopoulos}}]{Ellis:2011}
\bibinfo{author}{\bibfnamefont{J.}~\bibnamefont{Ellis}},
  \bibinfo{author}{\bibfnamefont{N.~E.} \bibnamefont{Mavromatos}},
  \bibnamefont{and} \bibinfo{author}{\bibfnamefont{D.~V.}
  \bibnamefont{Nanopoulos}}, \bibinfo{journal}{Int. J. Mod. Phys. A}
  \textbf{\bibinfo{volume}{26}}, \bibinfo{pages}{2243} (\bibinfo{year}{2011}).

\bibitem[{\citenamefont{Gambini and Pullin}(1999)}]{Gambini:1999}
\bibinfo{author}{\bibfnamefont{R.}~\bibnamefont{Gambini}} \bibnamefont{and}
  \bibinfo{author}{\bibfnamefont{J.}~\bibnamefont{Pullin}},
  \bibinfo{journal}{Phys. Rev. D} \textbf{\bibinfo{volume}{59}},
  \bibinfo{pages}{124021} (\bibinfo{year}{1999}).

\bibitem[{\citenamefont{Alfaro et~al.}(2002)\citenamefont{Alfaro,
  Morales-T\'ecotl, and Urrutia}}]{Alfaro:2002}
\bibinfo{author}{\bibfnamefont{J.}~\bibnamefont{Alfaro}},
  \bibinfo{author}{\bibfnamefont{H.~A.} \bibnamefont{Morales-T\'ecotl}},
  \bibnamefont{and} \bibinfo{author}{\bibfnamefont{L.~F.}
  \bibnamefont{Urrutia}}, \bibinfo{journal}{Phys. Rev. D}
  \textbf{\bibinfo{volume}{65}}, \bibinfo{pages}{103509}
  (\bibinfo{year}{2002}).

\bibitem[{\citenamefont{Ellis et~al.}(1992)\citenamefont{Ellis, Mavromatos, and
  Nanopoulos}}]{Ellis:1992}
\bibinfo{author}{\bibfnamefont{J.}~\bibnamefont{Ellis}},
  \bibinfo{author}{\bibfnamefont{N.}~\bibnamefont{Mavromatos}},
  \bibnamefont{and}
  \bibinfo{author}{\bibfnamefont{D.}~\bibnamefont{Nanopoulos}},
  \bibinfo{journal}{Phys. Lett. B} \textbf{\bibinfo{volume}{293}},
  \bibinfo{pages}{37 } (\bibinfo{year}{1992}).

\bibitem[{\citenamefont{Ellis et~al.}(1999)\citenamefont{Ellis, Mavromatos, and
  Nanopoulos}}]{Ellis:1999}
\bibinfo{author}{\bibfnamefont{J.}~\bibnamefont{Ellis}},
  \bibinfo{author}{\bibfnamefont{N.}~\bibnamefont{Mavromatos}},
  \bibnamefont{and}
  \bibinfo{author}{\bibfnamefont{D.}~\bibnamefont{Nanopoulos}},
  \bibinfo{journal}{Chaos, Solitons \& Fractals} \textbf{\bibinfo{volume}{10}},
  \bibinfo{pages}{345 } (\bibinfo{year}{1999}).

\bibitem[{\citenamefont{Amelino-Camelia}(2002)}]{Amelino-Camelia:2002}
\bibinfo{author}{\bibfnamefont{G.}~\bibnamefont{Amelino-Camelia}},
  \bibinfo{journal}{Int. J. Mod. Phys. D} \textbf{\bibinfo{volume}{11}},
  \bibinfo{pages}{35} (\bibinfo{year}{2002}).

\bibitem[{\citenamefont{Amelino-Camelia
  et~al.}(1998)\citenamefont{Amelino-Camelia, Ellis, Mavromatos, Nanopoulos,
  and Sarkar}}]{Amelino-Camelia:1998}
\bibinfo{author}{\bibfnamefont{G.}~\bibnamefont{Amelino-Camelia}},
  \bibinfo{author}{\bibfnamefont{J.}~\bibnamefont{Ellis}},
  \bibinfo{author}{\bibfnamefont{N.}~\bibnamefont{Mavromatos}},
  \bibinfo{author}{\bibfnamefont{D.~V.} \bibnamefont{Nanopoulos}},
  \bibnamefont{and} \bibinfo{author}{\bibfnamefont{S.}~\bibnamefont{Sarkar}},
  \bibinfo{journal}{Nature} \textbf{\bibinfo{volume}{393}},
  \bibinfo{pages}{763} (\bibinfo{year}{1998}).

\bibitem[{\citenamefont{Ellis and Mavromatos}(2013)}]{Ellis:2013}
\bibinfo{author}{\bibfnamefont{J.}~\bibnamefont{Ellis}} \bibnamefont{and}
  \bibinfo{author}{\bibfnamefont{N.~E.} \bibnamefont{Mavromatos}},
  \bibinfo{journal}{Astropart. Phys.} \textbf{\bibinfo{volume}{43}},
  \bibinfo{pages}{50} (\bibinfo{year}{2013}).

\bibitem[{\citenamefont{Chang et~al.}(2012)\citenamefont{Chang, Jiang, and
  Lin}}]{Chang:2012}
\bibinfo{author}{\bibfnamefont{Z.}~\bibnamefont{Chang}},
  \bibinfo{author}{\bibfnamefont{Y.}~\bibnamefont{Jiang}}, \bibnamefont{and}
  \bibinfo{author}{\bibfnamefont{H.-N.} \bibnamefont{Lin}},
  \bibinfo{journal}{Astropart. Phys.} \textbf{\bibinfo{volume}{36}},
  \bibinfo{pages}{47} (\bibinfo{year}{2012}).

\bibitem[{\citenamefont{Shao et~al.}(2010)\citenamefont{Shao, Xiao, and
  Ma}}]{Shao:2010}
\bibinfo{author}{\bibfnamefont{L.}~\bibnamefont{Shao}},
  \bibinfo{author}{\bibfnamefont{Z.}~\bibnamefont{Xiao}}, \bibnamefont{and}
  \bibinfo{author}{\bibfnamefont{B.-Q.} \bibnamefont{Ma}},
  \bibinfo{journal}{Astropart. Phys.} \textbf{\bibinfo{volume}{33}},
  \bibinfo{pages}{312} (\bibinfo{year}{2010}).

\bibitem[{\citenamefont{Zhang and Ma}(2014)}]{Zhang:2014wpb}
\bibinfo{author}{\bibfnamefont{S.}~\bibnamefont{Zhang}} \bibnamefont{and}
  \bibinfo{author}{\bibfnamefont{B.-Q.} \bibnamefont{Ma}},
  \bibinfo{journal}{Astropart. Phys.} \textbf{\bibinfo{volume}{61}},
  \bibinfo{pages}{108} (\bibinfo{year}{2014}).

\bibitem[{\citenamefont{Vasileiou et~al.}(2013)\citenamefont{Vasileiou,
  Jacholkowska, Piron, Bolmont, Couturier, Granot, Stecker, Cohen-Tanugi, and
  Longo}}]{Vasileiou:2013}
\bibinfo{author}{\bibfnamefont{V.}~\bibnamefont{Vasileiou}},
  \bibinfo{author}{\bibfnamefont{A.}~\bibnamefont{Jacholkowska}},
  \bibinfo{author}{\bibfnamefont{F.}~\bibnamefont{Piron}},
  \bibinfo{author}{\bibfnamefont{J.}~\bibnamefont{Bolmont}},
  \bibinfo{author}{\bibfnamefont{C.}~\bibnamefont{Couturier}},
  \bibinfo{author}{\bibfnamefont{J.}~\bibnamefont{Granot}},
  \bibinfo{author}{\bibfnamefont{F.~W.} \bibnamefont{Stecker}},
  \bibinfo{author}{\bibfnamefont{J.}~\bibnamefont{Cohen-Tanugi}},
  \bibnamefont{and} \bibinfo{author}{\bibfnamefont{F.}~\bibnamefont{Longo}},
  \bibinfo{journal}{Phys. Rev. D} \textbf{\bibinfo{volume}{87}},
  \bibinfo{pages}{122001} (\bibinfo{year}{2013}).

\bibitem[{\citenamefont{Abdo et~al.}(2009{\natexlab{a}})\citenamefont{Abdo,
  Ackermann, Ajello, Asano, Atwood, Axelsson, Baldini, Ballet, Barbiellini,
  Baring et~al.}}]{Abdo:2009}
\bibinfo{author}{\bibfnamefont{A.~A.} \bibnamefont{Abdo}},
  \bibinfo{author}{\bibfnamefont{M.}~\bibnamefont{Ackermann}},
  \bibinfo{author}{\bibfnamefont{M.}~\bibnamefont{Ajello}},
  \bibinfo{author}{\bibfnamefont{K.}~\bibnamefont{Asano}},
  \bibinfo{author}{\bibfnamefont{W.~B.} \bibnamefont{Atwood}},
  \bibinfo{author}{\bibfnamefont{M.}~\bibnamefont{Axelsson}},
  \bibinfo{author}{\bibfnamefont{L.}~\bibnamefont{Baldini}},
  \bibinfo{author}{\bibfnamefont{J.}~\bibnamefont{Ballet}},
  \bibinfo{author}{\bibfnamefont{G.}~\bibnamefont{Barbiellini}},
  \bibinfo{author}{\bibfnamefont{M.~G.} \bibnamefont{Baring}},
  \bibnamefont{et~al.}, \bibinfo{journal}{Nature}
  \textbf{\bibinfo{volume}{462}}, \bibinfo{pages}{331}
  (\bibinfo{year}{2009}{\natexlab{a}}).

\bibitem[{\citenamefont{Ellis et~al.}(2006)\citenamefont{Ellis, Mavromatos,
  Nanopoulos, Sakharov, and Sarkisyan}}]{Ellis:2006}
\bibinfo{author}{\bibfnamefont{J.}~\bibnamefont{Ellis}},
  \bibinfo{author}{\bibfnamefont{N.~E.} \bibnamefont{Mavromatos}},
  \bibinfo{author}{\bibfnamefont{D.~V.} \bibnamefont{Nanopoulos}},
  \bibinfo{author}{\bibfnamefont{A.~S.} \bibnamefont{Sakharov}},
  \bibnamefont{and} \bibinfo{author}{\bibfnamefont{E.~K.}
  \bibnamefont{Sarkisyan}}, \bibinfo{journal}{Astropart. Phys.}
  \textbf{\bibinfo{volume}{25}}, \bibinfo{pages}{402} (\bibinfo{year}{2006}).

\bibitem[{\citenamefont{Vasileiou et~al.}(2015)\citenamefont{Vasileiou, Granot,
  Piran, and Amelino-Camelia}}]{Vasileiou:2015}
\bibinfo{author}{\bibfnamefont{V.}~\bibnamefont{Vasileiou}},
  \bibinfo{author}{\bibfnamefont{J.}~\bibnamefont{Granot}},
  \bibinfo{author}{\bibfnamefont{T.}~\bibnamefont{Piran}}, \bibnamefont{and}
  \bibinfo{author}{\bibfnamefont{G.}~\bibnamefont{Amelino-Camelia}},
  \bibinfo{journal}{Nature Physics} \textbf{\bibinfo{volume}{11}},
  \bibinfo{pages}{344} (\bibinfo{year}{2015}).

\bibitem[{\citenamefont{Nemiroff et~al.}(2012)\citenamefont{Nemiroff, Connolly,
  Holmes, and Kostinski}}]{Nemiroff:2012}
\bibinfo{author}{\bibfnamefont{R.~J.} \bibnamefont{Nemiroff}},
  \bibinfo{author}{\bibfnamefont{R.}~\bibnamefont{Connolly}},
  \bibinfo{author}{\bibfnamefont{J.}~\bibnamefont{Holmes}}, \bibnamefont{and}
  \bibinfo{author}{\bibfnamefont{A.~B.} \bibnamefont{Kostinski}},
  \bibinfo{journal}{Phys. Rev. Lett.} \textbf{\bibinfo{volume}{108}},
  \bibinfo{eid}{231103} (\bibinfo{year}{2012}).

\bibitem[{\citenamefont{Ellis et~al.}(2003)\citenamefont{Ellis, Mavromatos,
  Nanopoulos, and Sakharov}}]{Ellis:2003}
\bibinfo{author}{\bibfnamefont{J.}~\bibnamefont{Ellis}},
  \bibinfo{author}{\bibfnamefont{N.}~\bibnamefont{Mavromatos}},
  \bibinfo{author}{\bibfnamefont{D.~V.} \bibnamefont{Nanopoulos}},
  \bibnamefont{and} \bibinfo{author}{\bibfnamefont{A.}~\bibnamefont{Sakharov}},
  \bibinfo{journal}{Astron. Astrophys} \textbf{\bibinfo{volume}{402}},
  \bibinfo{pages}{409} (\bibinfo{year}{2003}).

\bibitem[{\citenamefont{Boggs et~al.}(2004)\citenamefont{Boggs, Wunderer,
  Hurley, and Coburn}}]{Boggs:2004}
\bibinfo{author}{\bibfnamefont{S.~E.} \bibnamefont{Boggs}},
  \bibinfo{author}{\bibfnamefont{C.~B.} \bibnamefont{Wunderer}},
  \bibinfo{author}{\bibfnamefont{K.}~\bibnamefont{Hurley}}, \bibnamefont{and}
  \bibinfo{author}{\bibfnamefont{W.}~\bibnamefont{Coburn}},
  \bibinfo{journal}{Astrophys. J.} \textbf{\bibinfo{volume}{611}},
  \bibinfo{pages}{L77} (\bibinfo{year}{2004}).

\bibitem[{\citenamefont{Zhang and Ma}(2015)}]{Zhang:2015}
\bibinfo{author}{\bibfnamefont{S.}~\bibnamefont{Zhang}} \bibnamefont{and}
  \bibinfo{author}{\bibfnamefont{B.-Q.} \bibnamefont{Ma}},
  \bibinfo{journal}{Astropart. Phys.} \textbf{\bibinfo{volume}{61}},
  \bibinfo{pages}{108} (\bibinfo{year}{2015}).

\bibitem[{\citenamefont{Pan et~al.}(2015)\citenamefont{Pan, Gong, Cao, Gao, and
  Zhu}}]{Pan:2015}
\bibinfo{author}{\bibfnamefont{Y.}~\bibnamefont{Pan}},
  \bibinfo{author}{\bibfnamefont{Y.}~\bibnamefont{Gong}},
  \bibinfo{author}{\bibfnamefont{S.}~\bibnamefont{Cao}},
  \bibinfo{author}{\bibfnamefont{H.}~\bibnamefont{Gao}}, \bibnamefont{and}
  \bibinfo{author}{\bibfnamefont{Z.-H.} \bibnamefont{Zhu}},
  \bibinfo{journal}{Astrophys. J.} \textbf{\bibinfo{volume}{808}},
  \bibinfo{pages}{78} (\bibinfo{year}{2015}).

\bibitem[{\citenamefont{Xiao and Ma}(2009)}]{Xiao:2009}
\bibinfo{author}{\bibfnamefont{Z.}~\bibnamefont{Xiao}} \bibnamefont{and}
  \bibinfo{author}{\bibfnamefont{B.-Q.} \bibnamefont{Ma}},
  \bibinfo{journal}{Phys. Rev. D} \textbf{\bibinfo{volume}{80}},
  \bibinfo{pages}{116005} (\bibinfo{year}{2009}).

\bibitem[{\citenamefont{Kosteleck\'y and Mewes}(2013)}]{Kosteleck:2013}
\bibinfo{author}{\bibfnamefont{V.~A.} \bibnamefont{Kosteleck\'y}}
  \bibnamefont{and} \bibinfo{author}{\bibfnamefont{M.}~\bibnamefont{Mewes}},
  \bibinfo{journal}{Phys. Rev. Lett.} \textbf{\bibinfo{volume}{110}},
  \bibinfo{pages}{201601} (\bibinfo{year}{2013}).

\bibitem[{\citenamefont{Kahniashvili et~al.}(2006)\citenamefont{Kahniashvili,
  Gogoberidze, and Ratra}}]{Kahniashvili:2006}
\bibinfo{author}{\bibfnamefont{T.}~\bibnamefont{Kahniashvili}},
  \bibinfo{author}{\bibfnamefont{G.}~\bibnamefont{Gogoberidze}},
  \bibnamefont{and} \bibinfo{author}{\bibfnamefont{B.}~\bibnamefont{Ratra}},
  \bibinfo{journal}{Phys. Lett. B} \textbf{\bibinfo{volume}{643}},
  \bibinfo{pages}{81} (\bibinfo{year}{2006}).

\bibitem[{\citenamefont{Coleman and Glashow}(1999)}]{Coleman:1999}
\bibinfo{author}{\bibfnamefont{S.}~\bibnamefont{Coleman}} \bibnamefont{and}
  \bibinfo{author}{\bibfnamefont{S.~L.} \bibnamefont{Glashow}},
  \bibinfo{journal}{Phys. Rev. D} \textbf{\bibinfo{volume}{59}},
  \bibinfo{pages}{116008} (\bibinfo{year}{1999}).

\bibitem[{\citenamefont{Biesiada and Pi\'{o}rkowska}(2009)}]{Biesiada:2009}
\bibinfo{author}{\bibfnamefont{M.}~\bibnamefont{Biesiada}} \bibnamefont{and}
  \bibinfo{author}{\bibfnamefont{A.}~\bibnamefont{Pi\'{o}rkowska}},
  \bibinfo{journal}{Class. Quantum Grav.} \textbf{\bibinfo{volume}{26}},
  \bibinfo{pages}{125007} (\bibinfo{year}{2009}).

\bibitem[{\citenamefont{Gogberashvili et~al.}(2007)\citenamefont{Gogberashvili,
  Sakharov, and Sarkisyan}}]{Gogberashvili:2007}
\bibinfo{author}{\bibfnamefont{M.}~\bibnamefont{Gogberashvili}},
  \bibinfo{author}{\bibfnamefont{A.~S.} \bibnamefont{Sakharov}},
  \bibnamefont{and} \bibinfo{author}{\bibfnamefont{E.~K.}
  \bibnamefont{Sarkisyan}}, \bibinfo{journal}{Phys. Lett. B}
  \textbf{\bibinfo{volume}{644}}, \bibinfo{pages}{179} (\bibinfo{year}{2007}).

\bibitem[{\citenamefont{Schaefer}(1999)}]{Schaefer:1999}
\bibinfo{author}{\bibfnamefont{B.~E.} \bibnamefont{Schaefer}},
  \bibinfo{journal}{Phys. Rev. Lett.} \textbf{\bibinfo{volume}{82}},
  \bibinfo{pages}{4964} (\bibinfo{year}{1999}).

\bibitem[{\citenamefont{Kumar and Duran}(2009)}]{Kumar:2009ps}
\bibinfo{author}{\bibfnamefont{P.}~\bibnamefont{Kumar}} \bibnamefont{and}
  \bibinfo{author}{\bibfnamefont{R.~B.} \bibnamefont{Duran}},
  \bibinfo{journal}{Mon. Not. Roy. Astron. Soc.}
  \textbf{\bibinfo{volume}{400}}, \bibinfo{pages}{75} (\bibinfo{year}{2009}).

\bibitem[{\citenamefont{Abdo et~al.}(2009{\natexlab{b}})}]{Abdo:2009pg}
\bibinfo{author}{\bibfnamefont{A.~A.} \bibnamefont{Abdo}} \bibnamefont{et~al.},
  \bibinfo{journal}{Astrophys. J.} \textbf{\bibinfo{volume}{706}},
  \bibinfo{pages}{L138} (\bibinfo{year}{2009}{\natexlab{b}}).

\bibitem[{\citenamefont{Ackermann et~al.}(2010)}]{Ackermann:2010us}
\bibinfo{author}{\bibfnamefont{M.}~\bibnamefont{Ackermann}}
  \bibnamefont{et~al.}, \bibinfo{journal}{Astrophys. J.}
  \textbf{\bibinfo{volume}{716}}, \bibinfo{pages}{1178} (\bibinfo{year}{2010}).

\bibitem[{\citenamefont{Ackermann et~al.}(2011)}]{Ackermann:2011bu}
\bibinfo{author}{\bibfnamefont{M.}~\bibnamefont{Ackermann}}
  \bibnamefont{et~al.}, \bibinfo{journal}{Astrophys. J.}
  \textbf{\bibinfo{volume}{729}}, \bibinfo{pages}{114} (\bibinfo{year}{2011}).

\bibitem[{SWI()}]{SWIFT}
\emph{\bibinfo{title}{Swift data archive}},
  \bibinfo{howpublished}{http://swift.gsfc.nasa.gov/archive/grb\_table/}.

\bibitem[{\citenamefont{Kumar and Zhang}(2015)}]{Kumar:2015}
\bibinfo{author}{\bibfnamefont{P.}~\bibnamefont{Kumar}} \bibnamefont{and}
  \bibinfo{author}{\bibfnamefont{B.}~\bibnamefont{Zhang}},
  \bibinfo{journal}{Phys. Rep.} \textbf{\bibinfo{volume}{561}},
  \bibinfo{pages}{1 } (\bibinfo{year}{2015}).

\bibitem[{\citenamefont{{Kouveliotou} et~al.}(1993)\citenamefont{{Kouveliotou},
  {Meegan}, {Fishman}, {Bhat}, {Briggs}, {Koshut}, {Paciesas}, and
  {Pendleton}}}]{Kouveliotou:1993}
\bibinfo{author}{\bibfnamefont{C.}~\bibnamefont{{Kouveliotou}}},
  \bibinfo{author}{\bibfnamefont{C.~A.} \bibnamefont{{Meegan}}},
  \bibinfo{author}{\bibfnamefont{G.~J.} \bibnamefont{{Fishman}}},
  \bibinfo{author}{\bibfnamefont{N.~P.} \bibnamefont{{Bhat}}},
  \bibinfo{author}{\bibfnamefont{M.~S.} \bibnamefont{{Briggs}}},
  \bibinfo{author}{\bibfnamefont{T.~M.} \bibnamefont{{Koshut}}},
  \bibinfo{author}{\bibfnamefont{W.~S.} \bibnamefont{{Paciesas}}},
  \bibnamefont{and} \bibinfo{author}{\bibfnamefont{G.~N.}
  \bibnamefont{{Pendleton}}}, \bibinfo{journal}{Astrophys. J.}
  \textbf{\bibinfo{volume}{413}}, \bibinfo{pages}{L101} (\bibinfo{year}{1993}).

\bibitem[{\citenamefont{Piran}(1999)}]{Piran:1999}
\bibinfo{author}{\bibfnamefont{T.}~\bibnamefont{Piran}},
  \bibinfo{journal}{Phys. Rep.} \textbf{\bibinfo{volume}{314}},
  \bibinfo{pages}{575 } (\bibinfo{year}{1999}).

\bibitem[{\citenamefont{Band et~al.}(1993)\citenamefont{Band, Matteson, Ford,
  Schaefer, Palmer et~al.}}]{Band:1993}
\bibinfo{author}{\bibfnamefont{D.}~\bibnamefont{Band}},
  \bibinfo{author}{\bibfnamefont{J.}~\bibnamefont{Matteson}},
  \bibinfo{author}{\bibfnamefont{L.}~\bibnamefont{Ford}},
  \bibinfo{author}{\bibfnamefont{B.}~\bibnamefont{Schaefer}},
  \bibinfo{author}{\bibfnamefont{D.}~\bibnamefont{Palmer}},
  \bibnamefont{et~al.}, \bibinfo{journal}{Astrophys. J.}
  \textbf{\bibinfo{volume}{413}}, \bibinfo{pages}{281} (\bibinfo{year}{1993}).

\bibitem[{\citenamefont{Preece et~al.}(1998)\citenamefont{Preece, Pendleton,
  Briggs, Mallozzi, Paciesas, Band, Matteson, and Meegan}}]{Preece:1998}
\bibinfo{author}{\bibfnamefont{R.~D.} \bibnamefont{Preece}},
  \bibinfo{author}{\bibfnamefont{G.~N.} \bibnamefont{Pendleton}},
  \bibinfo{author}{\bibfnamefont{M.~S.} \bibnamefont{Briggs}},
  \bibinfo{author}{\bibfnamefont{R.~S.} \bibnamefont{Mallozzi}},
  \bibinfo{author}{\bibfnamefont{W.~S.} \bibnamefont{Paciesas}},
  \bibinfo{author}{\bibfnamefont{D.~L.} \bibnamefont{Band}},
  \bibinfo{author}{\bibfnamefont{J.~L.} \bibnamefont{Matteson}},
  \bibnamefont{and} \bibinfo{author}{\bibfnamefont{C.~A.}
  \bibnamefont{Meegan}}, \bibinfo{journal}{Astrophys. J.}
  \textbf{\bibinfo{volume}{496}}, \bibinfo{pages}{849} (\bibinfo{year}{1998}).

\bibitem[{\citenamefont{Lloyd and Petrosian}(2000)}]{Lloyd:2000}
\bibinfo{author}{\bibfnamefont{N.~M.} \bibnamefont{Lloyd}} \bibnamefont{and}
  \bibinfo{author}{\bibfnamefont{V.}~\bibnamefont{Petrosian}},
  \bibinfo{journal}{Astrophys. J.} \textbf{\bibinfo{volume}{543}},
  \bibinfo{pages}{722} (\bibinfo{year}{2000}).

\bibitem[{\citenamefont{Amelino-Camelia and
  Smolin}(2009)}]{Amelino-Camelia:2009}
\bibinfo{author}{\bibfnamefont{G.}~\bibnamefont{Amelino-Camelia}}
  \bibnamefont{and} \bibinfo{author}{\bibfnamefont{L.}~\bibnamefont{Smolin}},
  \bibinfo{journal}{Phys. Rev. D} \textbf{\bibinfo{volume}{80}},
  \bibinfo{pages}{084017} (\bibinfo{year}{2009}).

\bibitem[{\citenamefont{Jacob and Piran}(2008)}]{Jacob:2008}
\bibinfo{author}{\bibfnamefont{U.}~\bibnamefont{Jacob}} \bibnamefont{and}
  \bibinfo{author}{\bibfnamefont{T.}~\bibnamefont{Piran}}, \bibinfo{journal}{J.
  Cosmol. Astropart. Phys.} \textbf{\bibinfo{volume}{2008}},
  \bibinfo{pages}{031} (\bibinfo{year}{2008}).

\bibitem[{\citenamefont{Ellis et~al.}(2008{\natexlab{b}})\citenamefont{Ellis,
  Mavromatos, Nanopoulos, Sakharov, and Sarkisyan}}]{Ellis:2008}
\bibinfo{author}{\bibfnamefont{J.}~\bibnamefont{Ellis}},
  \bibinfo{author}{\bibfnamefont{N.~E.} \bibnamefont{Mavromatos}},
  \bibinfo{author}{\bibfnamefont{D.~V.} \bibnamefont{Nanopoulos}},
  \bibinfo{author}{\bibfnamefont{A.~S.} \bibnamefont{Sakharov}},
  \bibnamefont{and} \bibinfo{author}{\bibfnamefont{E.~K.~G.}
  \bibnamefont{Sarkisyan}}, \bibinfo{journal}{Astropart. Phys.}
  \textbf{\bibinfo{volume}{29}}, \bibinfo{pages}{158}
  (\bibinfo{year}{2008}{\natexlab{b}}).

\bibitem[{\citenamefont{Ade et~al.}(2015)}]{Ade:2015}
\bibinfo{author}{\bibfnamefont{P.}~\bibnamefont{Ade}} \bibnamefont{et~al.},
  \bibinfo{howpublished}{e-print arXiv:1502.01589} (\bibinfo{year}{2015}).

\bibitem[{\citenamefont{Rubtsov et~al.}(2012)\citenamefont{Rubtsov, Pshirkov,
  and Tinyakov}}]{Rubtsov:2012}
\bibinfo{author}{\bibfnamefont{G.~I.} \bibnamefont{Rubtsov}},
  \bibinfo{author}{\bibfnamefont{M.~S.} \bibnamefont{Pshirkov}},
  \bibnamefont{and} \bibinfo{author}{\bibfnamefont{P.~G.}
  \bibnamefont{Tinyakov}}, \bibinfo{journal}{Mon. Not. R. Astron. Soc.}
  \textbf{\bibinfo{volume}{421}}, \bibinfo{pages}{L14} (\bibinfo{year}{2012}).

\end{thebibliography}

\end{document}